\documentclass[
    letterpaper,
    twoside,
    prx, 
    aps,    
    twocolumn,
    reprint, 
    showpacs,
    showkeys, 
    superscriptaddress, 
    longbibliography,
]{revtex4-2} 

\usepackage{amsthm}
\usepackage{amsfonts}
\usepackage{amsmath}
\usepackage{amssymb}
\usepackage{booktabs}
\usepackage{mathtools}
\usepackage{multirow}
\usepackage{graphicx}
\usepackage[dvipsnames]{xcolor}
\usepackage{dcolumn}
\usepackage{bm}
\usepackage{hyperref} 
\hypersetup{
    linkcolor  = RedViolet,
    citecolor  = YellowOrange,
    urlcolor   = Aquamarine,
    colorlinks = true,
} 
\usepackage[capitalize]{cleveref} 
\usepackage{braket}
\usepackage{tabularx}

\newcommand{\QuTech}{\affiliation{QuTech, Delft University of Technology, P.O. Box 5046, 2600 GA Delft, The Netherlands}}
\newcommand{\DIAM}{\affiliation{
    Delft Institute of Applied Mathematics, Technische Universiteit Delft, 2628 CD Delft, The Netherlands}}
\newcommand{\Riverlane}{\affiliation{Riverlane, Cambridge, CB2 3BZ, United Kingdom}}

\newcommand{\ns}{\mathrm{ns}}
\newcommand{\us}{\mu\mathrm{s}}

\begin{document}

\title{Neural network decoder for near-term surface-code experiments}

\author{Boris~M.~Varbanov}
\email{Corresponding author: b.m.varbanov@tudelft.nl}
\QuTech{}

\author{Marc~Serra-Peralta}
\QuTech{}\DIAM{}

\author{David~Byfield}
\Riverlane{}

\author{Barbara~M.~Terhal}
\QuTech{} \DIAM{}

\date{\today}

\begin{abstract}
  Neural-network decoders can achieve a lower logical error rate compared to conventional decoders, like minimum-weight perfect matching, when decoding the surface code.
  Furthermore, these decoders require no prior information about the physical error rates, making them highly adaptable.
  In this study, we investigate the performance of such a decoder using both simulated and experimental data obtained from a transmon-qubit processor, focusing on small-distance surface codes.
  We first show that the neural network typically outperforms the matching decoder due to better handling errors leading to multiple correlated syndrome defects, such as \(Y\) errors.
  When applied to the experimental data of [Google Quantum AI, Nature 614, 676 (2023)], the neural network decoder achieves logical error rates approximately \(25\%\) lower than minimum-weight perfect matching, approaching the performance of a maximum-likelihood decoder.
  To demonstrate the flexibility of this decoder, we incorporate the soft information available in the analog readout of transmon qubits and evaluate the performance of this decoder in simulation using a symmetric Gaussian-noise model.
  Considering the soft information leads to an approximately \(10\%\) lower logical error rate, depending on the probability of a measurement error.
  The good logical performance, flexibility, and computational efficiency make neural network decoders well-suited for near-term demonstrations of quantum memories.
\end{abstract}
\maketitle

\section{Introduction}
Quantum computers are anticipated to outperform classical computers in solving specific problems, such as integer factorization~\cite{Shor97} and quantum simulation~\cite{Lloyd96}.
However, for a quantum computer to perform any meaningful computation, it has to be able to execute millions of operations, requiring error rates per operation lower than \(10^{-10}\)~\cite{Reiher17,Gidney19}. Despite a valiant experimental effort aimed at enhancing operational performance, state-of-the-art processors typically exhibit error rates per operation around \(10^{-3}\)~\cite{Barends14,Rol16,Barends19,Rol19,Negirneac20,Foxen20,Jurcevic21,Harty14,Hong19,Huang19}, which is far from what is needed to perform any useful computation.

Fortunately, quantum error correction (QEC) provides a means to reduce the error rates, albeit at the cost of additional overhead in the required physical qubits~\cite{Shor95,Knill98,Aharonov08,Gottesman14}.
Two-dimensional stabilizer codes~\cite{GottesmanPhD}, such as the surface codes~\cite{Kitaev03}, have emerged as a prominent approach to realizing fault-tolerant computation due to their modest connectivity requirements and high tolerance to errors~\cite{Dennis02,Fowler12,Raussendorf07}.
These codes encode the logical information into an array of physical qubits, referred to as data qubits.
Ancilla qubits are used to repeatedly measure parities of sets of neighboring data qubits.
Changes between consecutive measurement outcomes, which are typically referred to as syndrome defects, indicate that errors have occurred.
A classical decoder processes this information and aims at inferring the most likely correction.

The increased number of available qubits~\cite{Corcoles19,Arute19,Acharya23,Sundaresan23} and the higher fidelities of physical operations~\cite{Barends14,Rol16,Barends19,Rol19,Negirneac20,Foxen20,Jurcevic21,Harty14,Hong19,Huang19,Heinsoo18,Marques23,McEwen21,Miao22,Jeffrey14,Bultink16} in modern processors have enabled several experiments employing small-distance codes to demonstrate the capacity to detect and correct errors~\cite{Kelly15,Egan20,Abobeih22,RyanAnderson21,Marques22,Chen21,Andersen20,Krinner22,Zhao22,Acharya23,Sundaresan23,Ofek16,Grimm20,CampagneIbarcq20,Sivak23}. In a recent milestone experiment, the error rate per QEC~round of a surface-code logical qubit was reduced by increasing the code distance~\cite{Acharya23}, demonstrating the fundamental suppression achieved by~QEC\@.

The performance of the decoder directly influences the performance of a QEC code.
Minimum-weight perfect matching (MWPM) is a good decoding algorithm for the surface code, which is computationally efficient and, therefore, scalable~\cite{Dennis02,Fowler12d,Fowler15,Higgott23,Wu23}.
Its good performance is ensured under the assumption that the errors occurring in the experiment can be modeled as independent \(X\) and \(Z\) errors~\cite{Dennis02}.
This leads to the MWPM decoder performing worse than decoders based on belief propagation~\cite{Roffe20,Criger18,Higgott22,Caune23} or a (more computationally-expensive) approximate maximum-likelihood decoder based on tensor-network (TN) contraction~\cite{Bravyi14,Chubb21}.
A more practical concern is that a decoder relies on a physical error model to accurately infer the most likely correction.
Typically, this requires constructing an approximate model and a series of benchmarking experiments to extract the physical error rates.
While there are methods to estimate the physical error rates based on the measured defects~\cite{Spitz17,Chen21,Chen22,Acharya23}, they typically ignore non-conventional errors like crosstalk or leakage.
The presence of these errors can impact both the accuracy with which the physical error rates are estimated from the data and the performance of the decoder itself~\cite{Chen22}.

An alternative approach to decoding is based on using neural networks (NN) to infer the most likely correction given a set of measured defects~\cite{Torlai17,Krastanov17,Varsamopoulos18,Baireuther18,Chamberland18b,Baireuther19,Andreasson19,Ni20,Wagner20,Sheth20,Varsamopoulos20a,Varsamopoulos20b,Fitzek20,Sweke21,Meinerz22,Ueno22,Chamberland22,Overwater22,Gicev23,Zhang23,Egorov23}.
These decoders do not require any prior information about the error model and therefore alleviate the need to construct any error model, making them highly adaptable.
This flexibility comes at the cost of requiring a significant amount of data for training the network and optimizing the hyper-parameters to ensure that the optimal performance of the decoder is reached during training.
Despite the potential issues during the training, it has been shown that they can match and generally exceed the performance of MWPM decoders, in several cases achieving near-optimal performance~\cite{Baireuther18,Baireuther19}.
Depending on the NN architecture employed, these decoders can be scalable and run in real time~\cite{Overwater22,Chamberland22,Gicev23,Zhang23, Ni20}.
While decoders based on recurrent NNs are more computationally expensive, they enable the decoding of experiments performing a variable number of stabilizer measurement rounds~\cite{Baireuther18,Baireuther19,Varsamopoulos20a}, making them well-suited for decoding near-term memory~\cite{Baireuther18} and stability experiments~\cite{Gidney22}.

In this work, we assess the performance of a neural-network decoder using both simulated and experimental data.
Our work goes beyond~\cite{Baireuther18} and previous NN decoding works in applying and partially training a NN decoder for the first time on data from a surface-code experiment~\cite{Acharya23}, thus capturing realistic performance and showing the versatility of NN decoders. In addition, we go beyond~\cite{Baireuther18} in training the NN decoder for a distance-7 surface code and extract its exponential error suppression factor on simulated data. Thirdly, we show that our NN decoder can be trained with (simulated) soft measurement data and get a performance enhancement.

We begin by simulating the performance of a \(d=3\) surface code using a circuit-level noise model to show that the NN decoder outperforms MWPM by learning to deal with \(Y\) errors, as previous studies have suggested~\cite{Baireuther18}.

Next, we investigate the performance of the NN decoder when applied to data from a recent surface code experiment~\cite{Acharya23}.
Due to the limited volume of available experimental data, we train the NN decoder on simulated data generated using an error model based on the measured physical error rates.
However, we evaluate the decoder's performance on simulated and experimental data.
The NN decoder significantly outperforms MWPM when decoding simulated data and achieves a lower logical error rate for the \(d=5\) code than the constituent \(d=3\) codes.
When evaluated on experimental data, the NN decoder achieves a performance approaching that of a tensor-network decoder, which approximates a maximum-likelihood decoder.
However, contrary to the finding in~\cite{Acharya23}, the logical error rate observed in the \(d=5\) experiment is higher than the average of each of the \(d=3\) experiments, which we attribute to either a sub-optimal choice of hyper-parameters or the mismatch between the simulated data that the decoder was trained on and the experimental data.

To further explore the performance of NNs, we consider the continuous information available in the measurement outcomes of transmon qubits~\cite{Krantz19,Blais21}, typically referred to as soft information~\cite{Pattison21}.
By calculating the defect probabilities given the soft outcomes and providing them to the neural network during training and evaluation, we demonstrate that the soft decoder can achieve an approximately \(10\%\) lower logical error rate if the measurement error probability is sufficiently high.

\section{Background}
\subsection{The surface code}

\begin{figure}[htbp]
  \centering
  \includegraphics{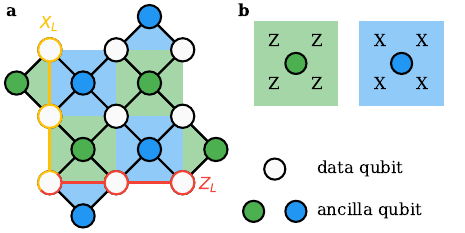}
  \caption{\label{fig:surface_code} \textbf{a} Schematic of a distance \(d=3\) surface-code logical qubit, where 9 data qubits (white circles) store the logical information. 8 ancilla qubits (blue and green circles) are used to measure the \(Z\)-type (green plaquettes) and \(X\)-type (blue plaquettes) stabilizers of the code. Examples of the \(X_{L}\) (yellow) and \(Z_{L}\) (red) logical operators of the code. \textbf{b} Illustration of the \(Z\)-type plaquette (left, green) and \(X\)-type (right, blue) plaquette corresponding to the \(ZZZZ\) and \(XXXX\) stabilizer operators measured by each ancilla qubit.}
\end{figure}

A (rotated) surface code encodes a single logical qubit into a two-dimensional array of \(n = d\times d\) physical qubits, referred to as data qubits, where \(d\) is the distance of the code.
The logical state of the qubit is determined by the stabilizers of the code, which are the weight-four or weight-two \(X\)-type (blue plaquettes) or \(Z\)-type (green plaquettes) Pauli operators, see~\cref{fig:surface_code}.
In addition to the stabilizers, the code is given by a pair of anti-commuting logical operators, \(X_L\) and \(Z_L\), which commute with the code stabilizers.
The stabilizers are typically measured indirectly with the help of \(n-1\) ancilla qubits.
To perform this measurement, each ancilla coherently interacts with its neighboring data qubits in a specific order~\cite{Tomita14}, after which the ancilla qubit is measured and reset.
The stabilizer measurement outcomes are typically referred to as the syndromes and hold information about the errors that have occurred.
The full circuits used to perform these measurements are shown in~\cref{fig:circuit}.
In particular, we use the circuits used in~\cite{Acharya23}, which feature several echo gates used for dynamical decoupling in the experiment, see~\cref{subsection:memory_exp} for additional details.

To characterize the performance of the code, we perform a series of logical memory experiments.
In each experiment, the physical qubits are prepared in an eigenstate of either the \(X_{L}\) (resp. \(Z_{L}\)) logical operator, after which \(N-1\) rounds of stabilizer measurements are executed.
The experiment is concluded by reading out each data qubit in the \(X\) (resp. \(Z\) basis), which also performs a logical \(X_L\) (resp. \(Z_L\)) measurement.
The goal of each experiment is to maintain the logical state for as many QEC~rounds as possible by using error correction, see~\cref{subsection:memory_exp} for more details.

The information about errors is contained in the stabilizer measurement outcome \(m_{r, a}\) of ancilla \(a\) at round \(r\).
The final data qubit measurements can also be used to infer a final set of outcomes \(m_{r=N, a}\) for either the \(X\)-type or \(Z\)-type stabilizers.
The defects \(d_{r, a} = m_{r, a} \oplus m_{r - 1, a}\) isolate the changes in \(m_{r, a}\) such that an error is signaled by an observation of one or more \(d_{r, a} = 1\).
The choice of initial state and the dynamical decoupling gates can also flip some of the measured \(m_{r, a}\), which is accounted for when calculating \(d_{r, a}\).
A decoder processes the observed \(d_{r,a}\) to infer a correction for the measured logical observable.
By repeating each experiment many times, we extract the probability of a logical error \(p_{L} \left( r \right)\) at QEC round \(r\), from which we calculate the logical fidelity \(F_{L} \left( r \right)  = 1 - 2p_{L} \left( r \right) \), which decays exponentially with the number of executed QEC~rounds.
We model this decay as \(F_{L}\left( r \right) = {\left( 1 - 2\varepsilon_{L} \right)}^{r - r_{0}}\), where \(\varepsilon_{L}\) is the logical error rate per QEC~round and \(r_{0}\) is a fitting constant.
When fitting the decay of \(F_{L}\left( r \right)\) to extract \(\varepsilon_{L}\), we start the fit at \(r=3\) to avoid any time-boundary effects that might impact this estimate.

\subsection{Error models}\label{sec:error_models}

To explore the performance of the NN decoder, we perform simulations using circuit-level Pauli-noise models. For most of our simulations, we consider a depolarizing circuit-level noise, which is defined as
\begin{enumerate}
  \item After each single-qubit gate or idling period, with a probability \(p/3\), we apply an error drawn from \(\{X, Y, Z\}\).
  \item After each two-qubit gate, with a probability \(p/15\), we apply an error drawn from \({\{I, X, Y, Z\}}^{\otimes 2} \setminus \{ II \}\).
  \item With a probability \(p\), we apply an \(X\) error before each measurement.
  \item With a probability \(p\), we apply an \(X\) error after each reset operation or after the qubits are first prepared at the start of an experiment.
\end{enumerate}

In some of our simulations, we consider noise models that are biased to have a higher or a lower probability of applying \(Y\) errors.
To construct this model, we define a Y-bias factor \(\eta\) and modify the standard depolarizing circuit-level noise model, as follows:
\begin{enumerate}
  \item After each single-qubit gate or idling period, there is a probability \(\eta p / (\eta + 2)\) to apply a \(Y\) error and a probability \(p / (\eta + 2)\) to apply an \(X\) or a \(Z\) error.
  \item After each two-qubit gate, there is a probability \(\eta p / (7\eta + 8)\) of applying an error drawn from \(\mathcal{P}_{B} =  \{ IY, XY, YI, YX, YY, YZ, ZY\}\) and a probability \( p / (7\eta + 8)\) of applying an error drawn from \({\{I, X, Y, Z\}}^{\otimes 2} \setminus  (\mathcal{P}_{B}  \cup \{ II\})\).
\end{enumerate}

This biased error model is a generalization of the depolarizing model.
In particular, choosing \(\eta = 1\) makes this noise model equivalent to the depolarizing one.
On the other hand, when \(\eta = 0\), the model leads to only \(X\) or \(Z\) errors applied after operations.
In the other limiting case, as \(\eta \to \infty\), the model applies only \(Y\) errors after idling periods and gates.
Given that the error probability is the same across all operations of the same type, we will refer to these error models as uniform circuit-level noise models.

Finally, we also perform simulations of the recent experiment conducted by Google Quantum AI, using the error model which they provided together with the experimental data~\cite{Acharya23}.
This is once again a circuit-level Pauli-noise model similar to the ones presented above, but the probability of a depolarizing error after each operation is based on the measured physical error rates.
We will refer to this model as the experimental circuit-level noise model.

We use \textit{stim}~\cite{Gidney21} to perform the stabilizer simulations. We have written a wrapper package that helps with constructing the circuit for each experiment, which is available in~\cite{VarbanovSurfaceSim23}. We use \textit{pymatching}~\cite{Higgott23} for the MWPM decoding. The weights used in the MWPM decoder are directly extracted from the sampled circuit using the built-in integration between \textit{stim} and \textit{pymatching}.

\subsection{Neural network architecture}
Here we describe the NN architecture that we employ in this work, which nearly exactly follows the one proposed in~\cite{Baireuther18, Baireuther19}.
Many NN decoders studied previously are based on feed-forward or convolutional NN architecture. These decoders can generally decode experiments running a fixed number of QEC rounds. Decoders based on recurrent NN architectures, on the other hand, can learn the temporal correlations in the data, allowing them to directly process experiments performing a variable number of QEC~rounds.
We have used the \textit{TensorFlow} library~\cite{TensorFlowLibrary} to implement the NN architecture, with the source code of the decoder available in~\cite{VarbanovQrennd23}, the parameters used for each training are listed in~\cref{tab:hyperparameters}, while the scripts that perform the training are available upon request.

\begin{figure}[htbp]
  \centering
  \includegraphics{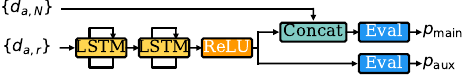}
  \caption{\label{fig:network} Schematic of the recurrent NN architecture used in this work, following the design proposed in~\cite{Baireuther19}. The inputs to the network are the set of defects \(\{d_{a, r}\}\), which are calculated from the measurement outcomes of each ancilla qubit \(a\) at QEC~round \(r=1,2,\ldots,N-1\), and the final defects \(\{d_{a, N}\}\), which are inferred from data qubit measurements. The time-invariant input \(\{d_{a, r}\}\) is provided to the recurrent part of the network, consisting of two stacked LSTM layers (yellow rectangles) and a ReLU activation layer (orange rectangle).  The recurrent output is then passed to the two heads of the decoder, which consist of an evaluation layer (blue rectangle) that predict a probability of a logical error. The lower head takes as input only the recurrent output and outputs a probability \(p_{\mathrm{aux}}\). The upper head, on the other hand, combines (teal rectangle) the recurrent output with \(\{d_{a, N}\}\) and outputs a probability \(p_{\mathrm{main}}\). Arrows indicate the flow of information through the network.}
\end{figure}

The NN architecture takes as input the defects \(d_{a,r}\) with \(r = 1, 2, \ldots, N\). The decoder solves a binary classification problem and determines whether a correction of the logical observable is required based on the observed defects.
In practice, the architecture is based on a two-headed network that makes two predictions \(p_{\mathrm{main}}\) and \(p_{\mathrm{aux}}\), which are used to improve the training of the network, see Fig.~\ref{fig:network}.
To train a decoder, a series of memory experiments are performed.
Since the logical qubit is prepared in a known logical state and measured at the end of each experiment, it is possible to extract the actual value \(p_{\mathrm{true}}\in \{0,1\}\) of whether a correction is required or not.
In particular, the cost function \(I\) that the network attempts to minimize during training is the weighted sum of the binary cross-entropies between each prediction and \(p_{\mathrm{true}}\), expressed as
\begin{equation*}
  I = H(p_{\mathrm{main}}, p_{\mathrm{true}}) + w_{a} H(p_{\mathrm{aux}}, p_{\mathrm{true}}),
\end{equation*}
where \(w_{a}\) is a weight that is typically chosen as \(w_{a} = 0.5\) in our runs, while
\begin{equation*}
  H(p_{i}, p_{j}) = - p_{i}\log p_{j} - (1 - p_{i})\log (1 - p_{j})
\end{equation*}
is the binary cross-entropy function. The choice behind this loss function is elaborated below.

\cref{fig:network}~schematically illustrates the architecture of the recurrent network.
The recurrent body of the neural network consists of two stacked long short-term memory (LSTM) layers.
Each LSTM layer is defined by a pair of internal memory states: a short-term memory, referred to as the hidden state, and a long-term memory, referred to as the cell state.
Here, we use the same internal states size \(N_L\) for both LSTM layers~\cite{Hochreiter97, LeCun2015}, with \(N_L = 64, 96, 128\) for surface codes of distance \(d=3,5,7\), unless otherwise specified.
The LSTM layers receive the defects for each QEC round as input, calculated from both the \(X\)-type and the \(Z\)-type stabilizer measurement outcomes.
The first LSTM layer outputs a hidden state for each QEC~round,  which is then provided as input to the second LSTM layer, which outputs only its final hidden state.
A rectified linear unit (ReLU) activation function is applied to the output of the second LSTM layer before being passed along to each of the two heads of the network.

The heads of the network are feed-forward evaluation networks consisting of a single hidden layer of size \(N_L\) using the ReLU activation function and an output layer using the sigmoid activation function, which maps the hidden layer output to a probability used for binary classification.
The output of the recurrent part of the network is directly passed to the lower head of the network, which uses this information to predict a probability \(p_{\mathrm{aux}}\) of a logical error.
The upper head also considers the defects inferred from the data qubit measurements, which are combined with the recurrent output and provided as input.
Therefore, unlike the lower head, the upper one uses the full information about the errors that have occurred when making its prediction \(p_{\mathrm{main}}\) of whether a logical error occurred.
Both \(p_{\mathrm{main}}\) and \(p_{\mathrm{aux}}\) are used when training the network, which helps the neural network to generalize to handle longer input sequences.
However, only \(p_{\mathrm{main}}\) is used when evaluating the performance of the decoder.
We provide additional details about the training procedure in~\cref{subsec:training} and list the hyper-parameters of the network in~\cref{tab:hyperparameters}.

\section{Results}

\subsection{Performance on circuit-level noise simulations}
We first demonstrate that the NN decoder can achieve a lower logical error rate than the MWPM decoder by learning error correlations between the defects, which are otherwise ignored by the MWPM~decoder.
We consider the \(Y\)-biased circuit-level noise model described previously, parameterized by the bias \(\eta\) towards \(Y\) errors and a probability \(p = 0.001\) of inserting an error after each operation.
We use this noise model to simulate the performance of a \(d=3\) surface-code quantum memory experiment in the \(Z\)-basis, initially preparing either \(\ket{0}^{\otimes n}\) or \(\ket{1}^{\otimes n}\).
To train the NN decoder, we generated datasets of \(r = 1, 5, \dots , 37\) QEC rounds, sampling \(5\times10^{5}\) shots for each round and initial state.
When evaluating the decoder's performance, we simulate the code performance over \(r = 10, 30, \dots , 290\) QEC~rounds and sample \(2\times10^{4}\) shots instead.

\begin{figure}[htbp]
  \centering
  \includegraphics{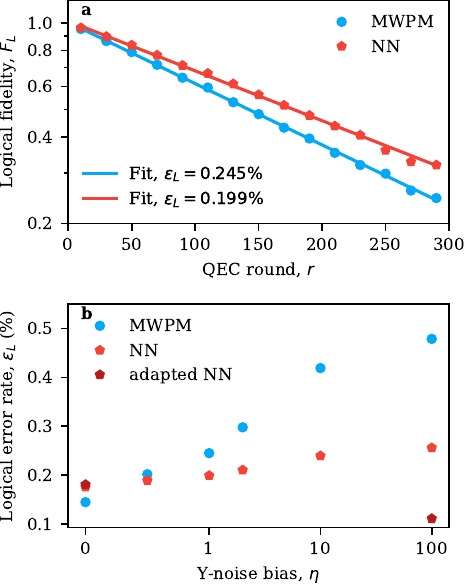}
  \caption{\label{fig:y_bias_performance} \textbf{a} Logical fidelity \(F_{L}\) as a function of the number of QEC rounds \(r\) for the MWPM (blue) and the NN decoders (red) using a uniform circuit-level depolarizing noise model. Each data point is averaged over \(4 \times 10^{4}\) shots. Solid lines show the fits to the data used to extract the logical error rate per round \(\varepsilon_{L}\). \textbf{b} The logical error rate \(\varepsilon_{L}\) as a function of the bias \(\eta\) towards \(Y\) errors for the MWPM~decoder (blue) and a NN decoder trained on simulated data using depolarizing noise (red), corresponding to \(\eta = 1\). The performance of an adapted NN decoder at a bias of \(\eta = 0\) or \(\eta = 100\) is shown in dark red. Each point is extracted from a fit of the decay of the logical fidelity over 300 QEC~rounds. The error bar is smaller than the marker size. The error bars are smaller than the marker sizes.}
\end{figure}

To benchmark the logical performance, we calculate the logical fidelity \(F_L\) at the end of each experiment.
Averaging \(F_L\) over each initial state, we fit the exponential decay of \(F_L\) with the number of QEC~rounds to extract the logical error rate per round \(\varepsilon_L\).
\cref{fig:y_bias_performance}~shows that the NN decoder maintains a constant \(\varepsilon_L\) when evaluated on datasets going up to \(300\) QEC rounds, demonstrating the ability of the decoder to generalize to significantly longer sequences than those used for training.
On the other hand, the NN decoder achieves about \(20\%\) lower \(\varepsilon_L\) compared to the MWPM decoder.
We then evaluate the trained NN decoder on simulated data using \(\eta \in \{0, 0.5, 1, 2, 10,  100\}\) and keep all other parameters the same without training any new neural networks, with the resulting error rates shown in~\cref{fig:y_bias_performance}\textbf{b}.
At \(\eta = 0\), corresponding to an error model leading to \(X\) and \(Z\) errors, the NN decoder displays a higher \(\varepsilon_L\) than the MWPM decoder.
For \(\eta \geq 0.5\), the NN decoder instead demonstrates a lower logical error, with the relative reduction increasing with the bias.
This demonstrates that the NN decoder can achieve a lower logical error rate by learning the correlations between the defects caused by \(Y\) errors, consistent with the results presented in~\cite{Baireuther18}.
The NN decoder can achieve an even lower logical error rate at a bias of \(\eta=100\) by being trained on a dataset generated using this bias (referred to as the adapted NN decoder in~\cref{fig:y_bias_performance}).
On the other hand, training a model for \(\eta = 0\) does not lead to any improvement in \(\varepsilon_L\) of the NN decoder, showing that the MWPM decoder is more optimal in this setting.

\subsection{Performance on experimental data}

\begin{figure}[htbp]
  \centering \includegraphics{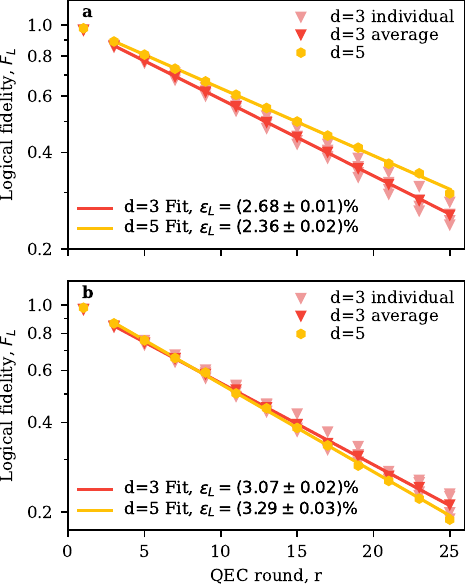}
  \caption{\label{fig:google_performance} Logical fidelity \(F_{L}\) as a function of the number of QEC~rounds \(r\) for the NN decoder evaluated on simulated data (shown in \textbf{a}) and on experimental data (shown in \textbf{b}). The average performance of the \(d=3\) surface code (red triangle), which is the average of the performance of each of the four constituent codes (bright red triangles), is compared to the \(d=5\) code (orange hexagons). Each data point is averaged over \(5 \times 10^{4}\) shots for both experiment and simulation. Solid lines show the fits to the data used to extract the logical error rate per round \(\varepsilon_{L}\). The error bars are smaller than the marker sizes.}
\end{figure}

Next, we evaluate the performance of the NN decoder on experimental data available from the recent experiment executed by Google Quantum AI~\cite{Acharya23}, where a 72-qubit quantum processor was used to implement a \(d=5\) surface code as well as the four \(d=3\) surface codes which use a subset of the qubits of the larger code.
The stabilizer measurement circuits used in that experiment are the same as those shown in~\cref{fig:circuit}.
For each distance-\(d\) surface code, the data qubits are prepared in several random bitstrings, followed by \(r=25\) rounds of stabilizer measurement, followed by a logical measurement, with experiments performed in both the \(X\)-basis and \(Z\)-basis.
The experiment demonstrated that the \(d=5\) surface code achieves a lower \(\varepsilon_L\) compared to the average of the four constituent \(d=3\) patches when using a tensor-network (TN) decoder, an approximation to a maximum-likelihood decoder.

We find that training a NN decoder to achieve good logical performance requires a large number of shots (approximately \(10^7\) in total or more) obtained from experiments preparing different initial states and running a different number of rounds.
As the amount of experimental data is too small to train the NN decoder (the total number of shots being \(6.5\times 10^{5}\)), we instead opt to simulate the experiments using the Pauli error model based on the measured error rates of each operation, available in~\cite{Acharya23}.
Keeping the same number of rounds and prepared state, we generate a total of \(2\times10^7\) shots for training the decoder for each \(d=3\) experiment and \(6\times10^7\) to train the decoder for the \(d=5\) experiment, see~\cref{tab:hyperparameters}.
While we train the network on simulated data, we still evaluate the decoder performance on both simulated and the experimental data, with the results shown in~\cref{fig:google_performance}\(\textbf{a}\) and~\cref{fig:google_performance}\(\textbf{b}\) respectively.
Both the training and evaluation data consist of \(r = 1, 3, \dots , 25\) rounds of QEC and consider the same initial states.
When evaluating the NN decoder on simulated data, we observe that the \(d=5\) code achieves a lower \(\varepsilon_L\) compared to the average of the \(d=3\) codes, see~\cref{fig:google_performance}\textbf{a}.
Evaluating the decoder on the experimental data leads to an approximately \(15\%\) (\(40\%\)) higher \(\varepsilon_L\) for the \(d=3\) (\(d=5\)) code, demonstrating that the approximate error model used in simulation fails to fully capture the errors in the experiment.
Furthermore, we observe that the \(d=5\) has a higher \(\varepsilon_L\) instead, see~\cref{fig:google_performance}\textbf{a}, contrary to what was demonstrated in~\cite{Acharya23} using a tensor-network decoder.

\begin{figure}[htpb]
  \centering
  \includegraphics{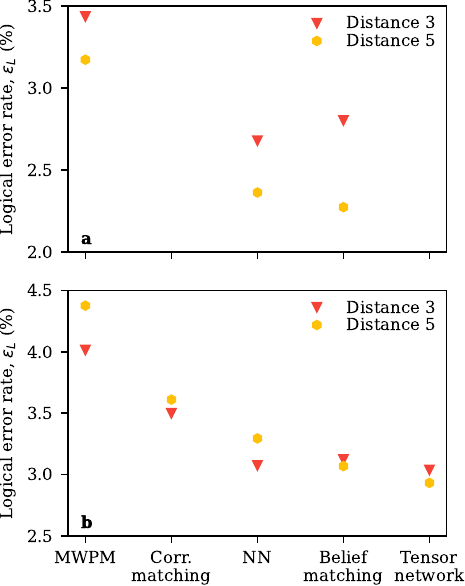}
  \caption{\label{fig:error_rate_comparison} The logical error rate per round \(\varepsilon_{L}\) for the \(d=3\) (red triangle) and \(d=5\) (orange hexagon) for several decoder implementations applied to either simulated data (shown in~\textbf{a}) or experimental data (shown in~\textbf{b}). These correspond (from left to right) to minimum-weight perfect matching (MWPM), a correlated modification of MWPM (Corr. MWPM)~\cite{Fowler13b}, our neural network (NN) decoder, belief matching (BM)~\cite{Higgott22}, and a tensor network  (TN) decoder, which approximates maximum-likelihood decoding.
    We did not run the corr. MWPM or TN decoder on the simulated data so fewer data points appear in~\textbf{a}. All logical error rates on the experimental data, except for the NN decoder, are taken from~\cite{Acharya23}.
    The error bars are smaller than the marker sizes.
  }
\end{figure}

In order to put the performance of the NN decoder in perspective, in~\cref{fig:error_rate_comparison}, we compare the logical performance of the NN decoder to the performance of several other decoders that were also implemented in~\cite{Acharya23}.
We perform this comparison both on simulated (see~\cref{fig:error_rate_comparison}\textbf{a}) and experimental (see~\cref{fig:error_rate_comparison}\textbf{b}) data.
We find that the NN decoder consistently outperforms the standard MWPM decoder in either case.
On the experimental dataset, the NN decoder performs equivalent to the TN decoder when decoding the \(d=3\) surface codes.
However, when decoding the \(d=5\) surface code experiment, the NN decoder displays a higher \(\varepsilon_L\) than the TN decoder and the computationally efficient belief-matching (BM) decoder~\cite{Higgott22}.
When evaluated on simulated data, the NN and BM decoders exhibit similar error rates, with the NN decoder again demonstrating better performance when decoding the \(d=3\) code but  worse when dealing with the \(d=5\) code.
The BM decoder we use for the simulated data is described in~\cite{Caune23} and uses the belief propagation implemented in~\cite{RoffeLDPCPythonTools22}.
The higher error rate of the NN decoder for the \(d=5\) code in both simulation and experiment can be related to the difficulty of optimizing the performance of the substantially larger NN model used (see~\cref{tab:hyperparameters} for the model hyper-parameters).
However, the discrepancy in the experiment can also be attributed to a mismatch between the simulated data used for training (based on an approximate error model) and the experimental data used for evaluation.
Compared to the \(d=3\) surface code data, the accumulation of qubit leakage can cause the \(d=5\) performance to degrade faster over the QEC~rounds~\cite{Acharya23}.
We expect that training on experimental data and a better hyper-parameter optimization to enable a NN performance comparable to state-of-the-art decoders like BM and TN while offering additional flexibility to the details of the noise model.
Compared to the TN decoder, both NN and BM can achieve similar logical performance while remaining significantly faster, and if their implementation is optimized, they can potentially be used to decode experiments in real time.

\subsection{Logical error rate suppression}\label{subsec:log_error_suppression}

\begin{figure}[htpb]
  \centering
  \includegraphics{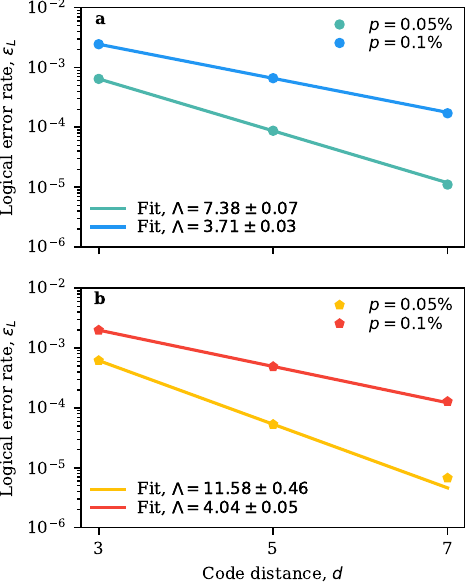}
  \caption{\label{fig:error_suppression} The logical error rate per round \(\varepsilon_{L}\) for surface codes of distance \(d=3, 5, 7\) for an MWPM decoder, shown in \textbf{a}, and our NN decoder, shown in \textbf{b}. This is evaluated on datasets using a uniform depolarizing circuit-level noise model with error probabilities of \(p=0.1\%\) (blue for the MWPM, red for the NN decoder) and \(p=0.05\%\) (teal for the MWPM, orange for the NN decoder). Solid lines show the fits to the data used to extract the logical error suppression factor \(\Lambda\). Each data point is extracted from a fit to the \(F_{L}\) as a function of QEC~rounds. The logical fidelities are extracted over \(10^{5}\) shots. The error bars are smaller than the marker sizes.
  }
\end{figure}

An exponential suppression of the logical error rate, assuming that the physical error rates are below `threshold', is vital for realizing a fault-tolerant quantum computer.
We explore the error suppression achieved when using the NN decoder.
We characterize the logical performance of \(d = 3, 5, 7\) surface codes simulated using a uniform depolarizing circuit-level noise model with an error probability of \(p=0.1\%\), close to the state-of-the-art physical error rates achieved in the experiment.
To train the NN decoder, we use data generated using this error probability.
We find that also training using a higher probability of \(p=0.2\%\) leads to a significantly lower logical error rate for the \(d=7\) code.
Furthermore, we evaluate the performance of the NN decoder on data simulated using \(p=0.05\%\), which is an example of the physical error rate needed to achieve practical sub-threshold scaling of the error rate.
For each distance \(d\) and error probability \(p\), we perform simulations of memory experiments in the \(Z\)-basis with varying numbers of QEC~rounds, going up to 600 rounds for the \(d=7\) code with an error rate of \(p=0.05\%\) to extract the logical error per round \(\varepsilon_L\).
The logical error rates obtained when using an MWPM decoder are shown in~\cref{fig:error_suppression}\textbf{a}, while those achieved by the NN decoder are shown in~\cref{fig:error_suppression}\textbf{b}.
If the physical error rate is below threshold, \(\varepsilon_L\) is expected to decay exponentially with the code distance \(d\), following \(\varepsilon_L\left(d\right) = C / \Lambda^{\left(d+1\right)/2}\), where \(\Lambda\) is the suppression factor and \(C\) is a fitting constant.
The data shows an apparent exponential suppression of the error rates by either decoder for the considered error rates, which we fit to extract the suppression factor \(\Lambda\), shown in~\cref{fig:error_suppression}.
In either case, the NN decoder achieves better logical performance compared to the MWPM decoder.
While for \(p=0.1\%\), the NN decoder achieves an approximately \(10\%\) higher \(\Lambda\), for \(p=0.05\%\), the more accurate NN decoder leads to an approximately \(60\%\) higher suppression factor instead.
The higher suppression factors \(\Lambda\) obtained from using better decoders significantly reduce the code distance required to achieve algorithmically-relevant logical error rates.
For example, for an error rate of \(p=0.05\%\), realizing \(\varepsilon_L \approx 10^{-10}\) would require a \(d=19\) surface code when using the MWPM decoder and \(d=15\) when using the NN decoder, corresponding to roughly \(40\%\) less physical qubits required.
However, whether the NN can continue to exhibit similar performance when decoding higher distance codes remains to be demonstrated.

\subsection{Decoding with soft information}
Measurements of physical qubits generally produce a continuous signal that is subsequently converted into declared binary outcomes by classical processing and thresholding. For example, transmon qubits are dispersively coupled to a dedicated readout resonator, which itself is connected to a readout feedline.
Readout is performed by applying a microwave pulse to the feedline, populating the readout resonator.
Due to a state-dependent shift of the resonator frequency, the outgoing signal is phase-shifted depending on whether the qubit is in the state \(\ket{0}\) or \(\ket{1}\).
This leads to a change in the real and imaginary components of the outgoing signal, which is experimentally measured.
This two-dimensional output can be transformed into a single continuous real variable and converted to a binary outcome by applying some threshold calibrated using a separate experiment~\cite{Blais21, Krantz19,Jeffrey14}.

While binary variables are convenient to work with and store, continuous measurement outcomes hold much more information about the state of the qubit, referred to as soft information.
It has been demonstrated that an MWPM-based decoder which considers the soft information of the individual measurements when decoding, offers higher thresholds and lower logical error rates than a hard decoder, which only considers the binary outcomes~\cite{Pattison21}.
To demonstrate the flexibility of machine-learning decoders, we consider providing the soft information available from readout when training and evaluating the NN decoder.

\begin{figure}[htpb]
  \centering
  \includegraphics{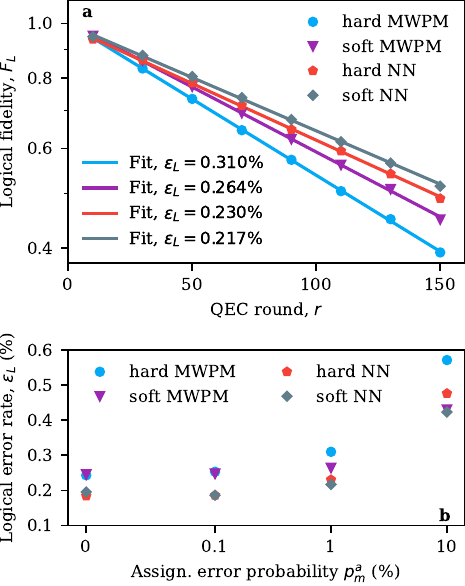}
  \caption{\label{fig:soft_performance} \textbf{a} The logical fidelity \(F_{L}\) as a function of the number of QEC rounds \(r\) for a hard and a soft version of the MWPM decoder (blue circles and purple triangles, respectively) and the NN decoder (red pentagons and gray diamonds, respectively). The soft decoders use the soft information by using the probability of observing defects in the case of the soft NN decoder or the likelihood of an assignment error~\cite{Pattison21} in the case of the soft MWPM decoder. The hard decoders use the defects obtained from the hard measurement outcomes. This performance is estimated on simulated data using a uniform depolarizing circuit-level noise model with an error probability \(p=0.1\%\). The soft outcome distributions are such that ancilla and data qubits have a probability of assignment errors of \(p_{m}^{a}=1\%\) and \(p_{m}^{d}=0.1\%\), respectively. Solid lines show the fits to the data used to extract the logical error rate per round \(\varepsilon_{L}\). Each data is averaged over \(10^{5}\) shots. \textbf{b} The extracted logical error rate \(\varepsilon_{L}\) for each of the three decoders as a function of the ancilla qubit assignment error probability \(p_{m}^{a}\), keeping \(p_{m}^{d}=0.1\%\) and \(p=0.1\%\). The error bars are smaller than the marker sizes.}
\end{figure}

In our simulations, measurements project the qubit into either \(\ket{0}\) or \(\ket{1}\).
A measurement outcome \(m_{r,q} = i\) of qubit \(q\) at round \(r\) corresponds to the ancilla qubit being in \(\ket{i}\) directly after the measurement.
Given \(m_{r,q} = i\), we model the soft outcome \(\tilde{m}_{r, q}\in \mathbb{R}\) to follow a Gaussian distribution \(\mathcal{N}_i\) with mean \(\mu_i\) and standard deviation \(\sigma\).
The soft outcome \(\tilde{m}_{r,q}\) can then be converted to a binary outcome \(\bar{m}_{r, q}\) by introducing a threshold \(t\), such that
\begin{equation*}
  \bar{m}_{r,a} =
  \begin{cases}
    0 & \text{if } \tilde{m}_{r, a} \leq t, \\
    1 & \text{otherwise.}
  \end{cases}
\end{equation*}
For the symmetric Gaussian distributions that we consider, this process leads to an assignment error probability \( P(\bar{m}_{r,q} = 0 \mid m_{r,q} = 1) = P(\bar{m}_{r,q} = 1 \mid m_{r,q} = 0) = p_{m}\).
This assignment error is {\emph{added}} to the errors considered in our circuit-level noise models, specifically the \(X\) error before each measurement that happens with a probability \(p\).
The assignment error probability can related to the signal-to-noise ratio \(\text{SNR}=\left| \mu_0 - \mu_1 \right| / 2\sigma\) as \(p_{m} = \frac{1}{2}\text{erfc}\left(\frac{{\rm SNR}}{\sqrt{2}}\right) \).
We fix \(\mu_0 = -1\) and \(\mu_1  = 1\) such that a given probability \(p_{m}\) fixes the standard deviation \(\sigma\) of the two distributions.

The most straightforward approach to incorporating the soft information into the NN decoder is to directly provide the soft measurement outcomes \(\tilde{m}_{r,q}\) as input during training and evaluation.
However, we find that doing this leads to an overall poor logical performance.
Instead, we estimate the probability of a defect \(P(d_{r,a} = 1 \mid \tilde{m}_{r,a}, \tilde{m}_{r-1,a})\), given the soft measurement outcomes of an ancilla qubit \(a\) in consecutive QEC rounds.
Given a soft outcome \(\tilde{m}_{r,q}\), the probability of the measured qubit `having being in the state' \(\ket{i}\) can be expressed as
\begin{equation*}
  P(i \mid \tilde{m}_{r,q}) = \frac{P(\tilde{m}_{r,q} \mid i)P(i)}{\sum_{j \in \{1, 2\}} P(\tilde{m}_{r,q} \mid j) P(j)}.
\end{equation*}
The soft outcomes follow a Gaussian distribution, that is, \(P(\tilde{m}_{r,q} \mid i) = \mathcal{N}_{i}(\tilde{m}_{r,q})\).
Finally, we make the simplifying assumption that the prior state probabilities \(P(i) = P(j) = \frac{1}{2}\), such that
\begin{equation*}
  P(i \mid \tilde{m}_{r,q}) = \frac{\mathcal{N}_{i}(\tilde{m}_{r,q})}{\sum_{j \in \{1, 2\}}\mathcal{N}_{j}(\tilde{m}_{r,q})}.
\end{equation*}
The probability of observing a defect can then be expressed as
\begin{equation*}
  \begin{aligned}
     & P(d_{r,a} = 1 \mid \tilde{m}_{r,a}, \tilde{m}_{r-1,a}) =                      \\
     & 1 - \sum_{i \in \{0,1\}}P(i \mid \tilde{m}_{r,a})P(i \mid \tilde{m}_{r-1,a}).
  \end{aligned}
\end{equation*}
The expression for the defect probability inferred from using the soft (final) data qubit measurement outcomes can be derived similarly.

To explore the performance of the soft NN decoder, we simulate the \(d=3\) surface-code memory experiment using a circuit-level noise model with an error rate per operation of \(p=0.1\%\).
We consider two separate assignment error probabilities \(p_{m}^{a}\) and \(p_{m}^{d}\) for ancilla qubit and data qubit measurements.
We motivate this choice by the fact that data qubits remain idling while the ancilla qubits are being measured.
A shorter measurement time can reduce the decoherence experienced by the data qubits but will typically lead to a higher \(p_{m}^{a}\).
The data qubit measurements at the end of the experiment, on the other hand, can be optimized to minimize \(p_{m}^{d}\).
Therefore, we focus on how a soft decoder can help with decoding when \(p_{m}^{a}\) is higher, similar to the discussion in~\cite{Pattison21}.
We train the NN decoder using datasets of \(r = 1, 5, \dots , 37\) QEC~rounds, sampling \(5\times10^{5}\) shots for each round and initial logical state.
When evaluating the performance, we use simulate \(r = 10, 30, \dots , 150\) QEC~rounds, sampling \(5\times10^{4}\) shots instead.

The results for \(p_{m}^{a}=1\%\) are shown in~\cref{fig:soft_performance}\textbf{a}. The hard NN decoder achieves an approximately \(20\%\) lower logical error rate than the hard MWPM decoder, consistent with the results shown in~\cref{fig:y_bias_performance}.
In comparison, the soft NN decoder leads to an approximately \(30\%\) lower logical error rate instead, demonstrating the ability of the decoder to adapt to the provided soft information.
Finally, we also compare the performance of these decoders to the soft MWPM decoder proposed in~\cite{Pattison21}.
This decoder encodes the soft information in the weights of the matching graph using the likelihood of an assignment error \(L_{r, a} = \mathcal{N}_{\neg i}(\tilde{m}_{r, a}) / \mathcal{N}_{i}(\tilde{m}_{r, a})\) given a soft outcome \(\tilde{m}_{r, a}\) that leads to a hard outcome of \(\bar{m}_{r, a} = i\).
We observe that using the soft MWPM decoder reduces the logical error rate by approximately \(15\%\) relative to the hard MWPM decoder, indicating that the soft NN decoder is not optimally using the available soft information.
In~\cref{fig:soft_performance}\textbf{b} the logical error rate \(\varepsilon_{L}\) of the three decoders is shown for \(p_{m}^{a} \in \{0, 0.1\%, 1\%, 10\%\}\), where both NN decoders are trained at the corresponding \(p_{m}^{a}\).
For low \(p_{m}^{a}\), the performance of the soft NN decoder is essentially equivalent to the hard NN decoder, with a moderate reduction in \(\varepsilon_L\) achieved for \(p_{m}^{a} \geq 1\%\).
We observe that the performance of the soft MWPM decoder becomes closer to that of the soft NN decoder as \(p_{m}^{a}\) increases, demonstrating that the probability of defects is likely not the optimal way to provide the soft information to the decoder.
Another downside of this representation is that for a high assignment error probability \(p_{m}^{a} \geq 20\%\), the probability of observing a defect is close to \(50\%\), which also impacts the training and leads the soft NN decoder to exhibit a higher logical error rate compared to the hard one (not shown in~\cref{fig:soft_performance}).
Finding a more optimal representation of the soft information that can be provided to the NN decoder and optimizing its performance remain open questions.

\section{Discussion}
We now discuss in more detail the performance of the NN decoder on the experimental data.
Unfortunately, we only use simulated data to train the NN decoder throughout this work.
These simulations use approximate Pauli-noise models that account for the most significant error mechanisms in the experiment, such as decoherence and readout errors.
However, they do not include several important error sources present in the actual experiments, such as leakage, crosstalk, and stray interactions.
The exclusion of these error mechanisms leads to the Pauli-noise models underpredicting the logical error rate compared to the rates observed in the experiment, as observed in~\cref{fig:google_performance}.
Furthermore, it was shown that the \(d=5\) code is more sensitive to errors like leakage and crosstalk, which can lead to a more significant deviation relative to simulations of the \(d=3\) codes~\cite{Acharya23}.
Despite using these approximate models for training, when evaluating the NN decoder on experimental data, we observe that it outperforms MWPM and can achieve logical error rates comparable to those obtained using maximum-likelihood decoding, which is approximated by the TN decoder.
The TN decoder requires information about the error probabilities, what defects they lead to, and their corresponding corrections, which can be encoded into a hypergraph, where the nodes correspond to defects and the hyperedges represent errors.
Importantly, this hypergraph also does not explicitly include hyperedges corresponding to non-conventional errors, such as leakage or crosstalk.
We expect that training on experimental data and optimizing the hyper-parameters of the network will enable it to match the performance of the TN decoder closely and potentially exceed it by learning about errors not included in the hypergraph.

Despite the large volume of training data required to achieve good performance, we don't expect that generating sufficient experimental data for training will be an issue.
Assuming that the QEC~round duration is \(1~\us\) and that it takes \(200~\ns\) to reset all qubits between subsequent runs, we estimate that it would take approximately three minutes to generate the datasets with \(10^{7}\) shots running \(r=1, 5, \ldots, 37\) rounds of QEC that were used for training the \(d=3,5,7\) surface codes, see~\cref{tab:hyperparameters}.

The soft NN decoder used in this work achieves only a moderate performance increase compared to the hard NN decoder. Furthermore, it uses the available soft information less optimally than the soft MWPM decoder. An alternative approach to incorporating the soft information into the decoder is to estimate the likelihood of assignment errors \(L_{r, a}\) used by the soft MWPM decoder and to provide them as input to the NN decoder together with the (hard) defects \(d_{r, a}\) that were measured.
In addition to the representation of the input data, it is an open question whether using a soft NN decoder will be useful in practice, where assignment error rates are typically low.
Specifically, it would be interesting to see if using a soft NN decoder will enable using a shorter measurement time that might lead to a higher assignment error rate but maximize the logical performance overall, as discussed in~\cite{Pattison21}.
The symmetric Gaussian distributions of the continuous measurement outcomes we consider here are only very simple approximations of the distributions seen in experiments, and in our modeling we could adapt these.
In particular, the relaxation that the qubit experiences during the readout leads to an asymmetry between the distributions and a generally higher probability of an assignment error when the qubit was prepared in \(\ket{1}\).
Furthermore, the continuous outcomes observed in the experiment can also contain information about leakage~\cite{Heinsoo18,Sank16,Khezri22} or correlations with other measurements.
Therefore, it will be essential to investigate and optimize the performance of the soft decoders using experimental data.

Finally, we outline some possible directions for future research necessary to use these decoders for decoding large-distance experiments.
Decoders based on feedforward and convolutional architectures have been shown to achieve low-latency decoding, making them a possible candidate for being used in real time~\cite{Overwater22,Chamberland22,Gicev23,Zhang23}.
On the other hand, recurrent networks generally have a larger number of parameters and carry out more complex operations when processing the data.
However, recurrent NN decoders have been shown to achieve higher accuracy and be more easily trainable than other architectures, especially when considering realistic noise models~\cite{Varsamopoulos20a}.
Therefore, whether hardware implementations of recurrent NN decoders can be used for real-time decoding is an open question.
In addition to the latency, the scalability of NN decoders is an open question.
Decoding higher-distance codes will require larger neural networks and larger training datasets, which will most likely be more challenging to train, given that approaches based on machine learning generally struggle when the dimension of the input becomes very large.
Practically, one might be interested in whether the NN decoder can be trained and used to decode some finite code distance, which is expected to lead to algorithmically-relevant logical error rates given the processor's performance.
Alternatively, there exist approaches that enable scalable NN decoders.
These are typically based on convolutional neural networks that learn to infer and correct the physical errors that have occurred while a secondary global decoder handles any possibly remaining errors~\cite{Chamberland22, Gicev23}, but a purely convolutional NN method has been explored as well~\cite{Ni20}.
The recurrent NN decoder used in this work is not scalable, and adapting it to work with larger code distances and using it to decode through logical operations is another open research venue.

Lastly, while preparing this manuscript, we became aware of a similar work~\cite{Lange23} that explores the performance of a graph neural network decoder on data from the repetition code experiment that was also done in~\cite{Acharya23}. More recently, Ref.~\cite{Bausch23} developed a transformer-based recurrent NN decoder and applied it to the surface code experiments~\cite{Acharya23} considered in this work, achieving a lower logical error rate than the TN decoder and demonstrating that the performance of such a decoder can be further improved by considering leakage in addition to the soft information.

\section*{Acknowledgments}
We are grateful to Earl Campbell for insightful discussions and for comments on the manuscript.
We also thank Laura Caune for implementing the belief-matching decoder that we have used in this work.
B.~M.~V. and B.~M.~T.~are supported by QuTech NWO funding 2020-2024 – Part I “Fundamental Research” with project number 601.QT.001-1. 
B.~M.~T and M.~S.-P. thank the OpenSuperQPlus100 project (no.~101113946) of the EU Flagship on Quantum Technology (HORIZON-CL4-2022-QUANTUM-01-SGA) for support. 

\section{Data and Software Availability}
The data and software that support the plots presented in this figure are available at~\cite{VarbanovFigureData23}.
The raw simulated data and the scripts used for training and decoding this data are available upon reasonable request.

\section{Appendix}
\subsection{Quantum memory experiments}\label{subsection:memory_exp}

\begin{figure*}[htpb]
  \centering
  \includegraphics[width=\textwidth]{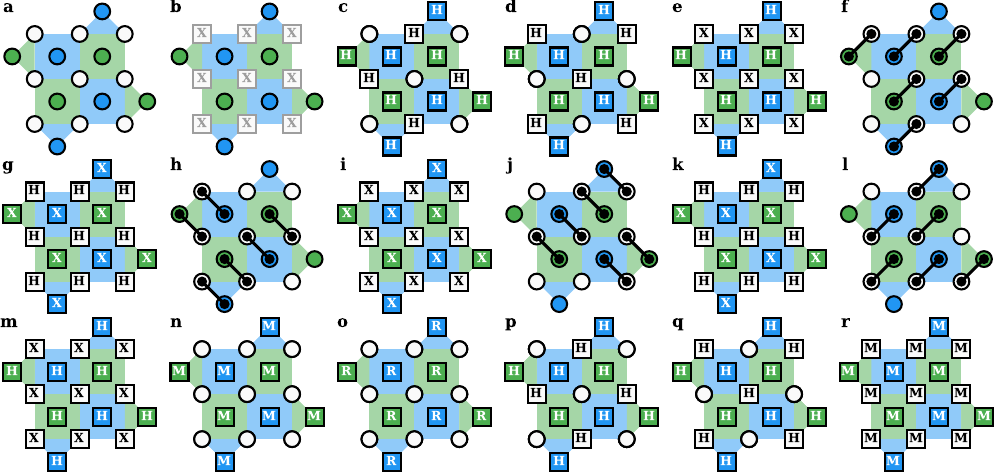}
  \caption{\label{fig:circuit} Schematic of the circuits used in the quantum memory experiments for a \(d=3\) surface code. \textbf{a-d} are used to initialize the logical state at the start of each experiment. The qubits are first prepared in the ground state (\textbf{a}), after which a set of conditional \(X\) gates (gray) are used to prepare the data qubits in a bit-string state (\textbf{b}). Afterward, a set of \(H\) (Hadamard) gates transform this into an eigenstate of the \(X\)-type (\textbf{c}) or \(Z\)-type (\textbf{d}) stabilizers. \textbf{e-o} show the circuits used to measure the stabilizers. The ancilla qubits are first placed in a superposition by a set of \(H\) gates (\textbf{c} or \textbf{d} in the first round, \textbf{e} otherwise). The parity of the neighboring data qubits is then mapped using four CZ gates (\textbf{f}, \textbf{h}, \textbf{j}, and \textbf{l}). The order of the gates used to measure the \(X\)- and \(Z\)-type stabilizers are chosen to avoid any “hook” errors propagating to a logical error. Two layers of \(H\) gates are applied to the data qubits (\textbf{g} and \textbf{k}) to measure the parity in the \(X\)-basis. In the middle of this sequence, \(X\) gates are applied to all qubits (\textbf{i}) for dynamical decoupling. Finally, the ancilla qubits are rotated back (\textbf{m}) using a set of \(H\) gates, measured (\textbf{n}, denoted by \(M\)) and reset (\textbf{o}, denoted by \(R\)). Several \(X\) gates are applied to the data qubit throughout this sequence for dynamical decoupling. In the final round, all data qubits are measured \textbf{p-r}, which is also a logical measurement. Some of the data qubits are rotated depending on whether the experiment is done in the \(X\) (\textbf{p}) or \(Z\) (\textbf{q}) logical basis. This step replaces \textbf{m} in the final round. Afterward, all qubits are measured simultaneously (\textbf{r}), replacing \textbf{n} in the final round. Data qubits are denoted with white circles, while ancilla qubits are illustrated as blue and green circles. For the definition of the plaquettes, see~\cref{fig:surface_code}.
    The circuits we run follow the ones used in~\cite{Acharya23}.
  }
\end{figure*}

To characterize the logical performance of a surface code, we look at its ability to maintain an initial logical state as a function of the number of QEC~rounds, commonly referred to as a quantum memory experiment.
The circuits used to perform these experiments are illustrated in~\cref{fig:circuit} and follow the ones used in the recent \(d=5\) surface code experiment done by Google Quantum AI~\cite{Acharya23}.
Removing some of the Hadamard gates when compiling the stabilizer measurement circuits leads to each ancilla qubit measuring the \(ZXXZ\) operator instead of the standard \(XXXX\) and \(ZZZZ\) stabilizers of the surface code.
Implementing this \(ZXXZ\) variant of the surface code symmetrizes the logical error rates between experiments done in the logical \(X\)-basis or \(Z\)-basis~\cite{Acharya23}.
Despite this modification, we use notations associated with the traditional stabilizers measured by the surface code.

Each experiment begins by preparing a given logical state, performed by the circuits in~\cref{fig:circuit}~\textbf{a-d}.
The data qubits are first initialized in the ground state and then prepared in either \(\ket{0}\) or \(\ket{1}\) by a layer of conditional \(X\) gates.
A subset of the data qubits is then rotated and transforms the initial state into an eigenstate of the \(X\)- or \(Z\)-type stabilizers.
The parity of the initial bistring state determines whether \(\ket{0}_L\) or \(\ket{1}_L\) (\(\ket{+}_L\) or \(\ket{-}_L\)) is prepared if the experiment is done in the \(Z\)-basis (\(X\)-basis).
In simulation, we prepare either \(\ket{0}^{\otimes n}\) or \(\ket{1}^{\otimes n}\) when using uniform circuit-level noise models.
In the experiment, several random bitstring states are used in order to symmetrize the impact of amplitude damping~\cite{Acharya23}.

The prepared logical state is then maintained over a total of \(r \in \{1, 2, \ldots, N-1 \} \) QEC rounds, with the circuit given by~\cref{fig:circuit}~\textbf{e-o}.
The first QEC~round then projects this initial state into a simultaneous eigenstate of both the \(X\)- or \(Z\)-type stabilizers.
Each cycle involves a series of four interactions between each ancilla qubit and its neighboring data qubits, which map the \(X\) or \(Z\) parity onto the state of the ancilla qubit.
The order in which these two-qubit operations are executed is carefully chosen to minimize the impact of errors occurring during the execution of the circuit~\cite{Tomita14}.
At the end of each QEC round, all of the ancilla qubits are measured and reset.
The stabilizer measurement circuits also contain several \(X\) gates on either the data or ancilla qubits, which dynamically decouple the qubits in the experiment~\cite{Acharya23}.
Naturally, these gates do not improve the logical performance for the simulations using approximate Pauli-error models that we consider here.
In the final QEC~round, the data qubits rotated during the state preparation are rotated back and measured in the \(Z\)-basis together with the ancilla qubits, illustrated in~\cref{fig:circuit}~\textbf{p-r}.
The data qubit measurement outcomes are then used to calculate the value of the \(X_{L}\) or \(Z_{L}\) logical observable as well as to infer a final set of \(X\)- or \(Z\)-type stabilizer measurement outcomes.

\subsection{Decoder training and evaluation}\label{subsec:training}

\begin{table*}[ht!]
  \centering
  \begin{tabular*}{0.8\textwidth}{@{\extracolsep{\fill}} ccccccc }
    \toprule
    \textbf{Distance}                                                &
    \textbf{Shots}                                                   &
    \textbf{Rounds}                                                  &
    \textbf{Dim. \(N_{L}\)}                                          &
    \textbf{\begin{tabular}[c]{@{}c@{}}Learning\\ rate\end{tabular}} &
    \textbf{\begin{tabular}[c]{@{}c@{}}Batch\\ size\end{tabular}}    &
    \textbf{\begin{tabular}[c]{@{}c@{}}Dropout\\ rate\end{tabular}}                                                                             \\ \midrule
    \multicolumn{7}{c}{Experimental circuit-level noise}                                                                                        \\ \midrule
    3                                                                & \(2\times10^{7}\) & [1, 25, 2] & 64  & \(5\times10^{-4}\)   & 64  & 5\%  \\ \midrule
    5                                                                & \(6\times10^{7}\) & [1, 25, 2] & 253 & \(5 \times 10^{-4}\) & 256 & 5\%  \\ \midrule
    \multicolumn{7}{c}{Uniform circuit-level noise}                                                                                             \\ \midrule
    3                                                                & \(10^{7}\)        & [1, 37, 4] & 64  & \(10^{-3}\)          & 256 & 20\% \\ \midrule
    5                                                                & \(10^{7}\)        & [1, 37, 4] & 96  & \(10^{-3}\)          & 256 & 20\% \\ \midrule
    7                                                                & \(10^{7}\)        & [1, 37, 4] & 128 & \(10^{-3}\)          & 256 & 20\% \\ \bottomrule
  \end{tabular*}%
  \caption{The hyper-parameters used for training the NN decoders. Different parameters are used for simulations based on the uniform circuit-level noise model and the experimental circuit-level noise, which models the experiments done in~\cite{Acharya23}. The internal state size of the network layers \(N_{L}\) is chosen to scale with the code distance \(d\). The QEC~round parameters \([i,j,k]\) for each dataset refer to performing experiments starting with \(i\) QEC~rounds and going up to \(j\) rounds in steps of \(k\). The total number of shots used for training is given, which is equally divided over the QEC~rounds and prepared states (not shown in the table). The learning rate, batch size, and dropout rate are the hyper-parameters we tune to help the network to train.}\label{tab:hyperparameters}
\end{table*}

Here we provide additional details about how we train the NN decoder and the hyper-parameters we use.
We use the Adam optimizer typically with a learning rate of \(10^{-3}\) or \(5 \times 10^{-4}\) for training.
In addition, we apply dropout after the hidden layer of the feed-forward network of each head and, in some cases, after the second LSTM layer with a dropout rate of either \(20\%\) or \(5\%\) to avoid over-fitting and assist with the generalization of the network.
We use a batch size of 256 or 64, which we found to lead to a smoother minimization of the loss.
After each training epoch, we evaluate the loss of the network on a separate dataset that considers the same number of QEC~rounds and prepared states as the training dataset but samples fewer shots for each experiment.
After each epoch, we save the networks' weights if a lower loss has been achieved.
Furthermore, we use early stopping to end the training if the loss has not decreased over the last 20 epochs to reduce the time it takes to train each model.
We have observed that not using early-stopping and leaving the training to continue does not typically lead the network to reach a lower loss eventually.
For some datasets, we lower the learning rate after the initial training has stopped early and train the network once more to achieve better performance.
The hyper-parameters we have used for training each network and the parameters of the training datasets used are presented in~\cref{tab:hyperparameters}.

The NN architecture we employ in this work uses two stacked LSTM layers to process the recurrent input~\cite{Baireuther19}.
We observe poor logical performance for a \(d=3\) surface code when using only a single LSTM layer.
On the other hand,  we see no significant improvement in the logical error rate when using four layers instead, motivating the choice to use only two.
This network architecture also performs well when decoding \(d=5\) and \(d=7\) surface code experiments.
However, we expect that a deeper recurrent network might improve the logical error rates when decoding larger-distance codes or when training on and decoding experimental data.
We have also practically observed that training the NN decoder for larger distances is more challenging, especially if the physical error rates are small.
Training the neural network on a dataset with a higher physical error rate (in addition to data using the same error rate as the evaluation dataset) can also improve the performance of the decoder, as we also discussed in~\cref{subsec:log_error_suppression}.

The training of our neural networks was performed on the DelftBlue supercomputer~\cite{DHPC2022} and was carried out on an NVIDIA Tesla V100S GPU\@.
Once trained, the decoder takes approximately 0.7 seconds per QEC~round for a \(d=3\) surface code (corresponding to an internal state size of \(N_{L} = 64\)) using a batch size of 50000 shots on an Intel(R) Core(TM) i7-8850H CPU @ 2.60GHz.  
For a \(d=5\) surface code (\(N_{L} = 96\)),  it takes about 0.8 seconds per round, while for a \(d=7\) surface code (\(N_{L} = 128\)), it takes about 1.1 seconds per round, using the same batch size of 50000 shots.
We note that using smaller batch sizes leads to a higher overall runtime due to parallelism when the network processes the inputs.
Therefore, larger batch sizes are preferable as long as they fit into the memory.
Each runtime was extracted by decoding simulated datasets running \(r=10, 30, \ldots, 290\) rounds of QEC and averaging the runtime per QEC~round over all the datasets.

\end{document}